\documentclass[11pt]{article}
\usepackage{amsmath}
\usepackage{booktabs}
\usepackage[margin=1in]{geometry}
\usepackage{graphicx}
\usepackage{rotating}
\usepackage{setspace}

\usepackage{url} %

\newcommand{\bu}{\ensuremath{\mathbf{u}}}
\newcommand{\bV}{\ensuremath{\mathbf{V}}}

\newcommand{\bb}{\ensuremath{\mathbf{b}}}
\newcommand{\bbeta}{\ensuremath{\boldsymbol{\beta}}}

\DeclareMathOperator{\Cov}{Cov}

\newcommand\indep{\protect\mathpalette{\protect\independenT}{\perp}}
\def\independenT#1#2{\mathrel{\rlap{$#1#2$}\mkern2mu{#1#2}}}

\usepackage{natbib}
\bibpunct{[}{]}{;}{a}{,}{;}

\begin{document}

\newlength{\oparindent}
\setlength{\oparindent}{\parindent}
\setlength{\parindent}{0pt}


%

Statistical testing of shared genetic control for potentially related
traits

\vspace{1cm}

Chris Wallace

(Corresponding author)
 \vspace{1cm}

Department of Medical Genetics\\
JDRF/Wellcome Trust Diabetes and Inflammation Laboratory\\
NIHR Cambridge Biomedical Research Centre\\
Cambridge Institute for Medical Research\\
University of Cambridge\\
Wellcome Trust/MRC Building\\
Cambridge\\
CB2 0XY\\ UK\\
+44 (0)1223 761093\\
  \url{chris.wallace@cimr.cam.ac.uk}

\newpage

\setlength{\parindent}{\oparindent}

\begin{abstract}
  Integration of data from genome-wide single nucleotide polymorphism
  (SNP) association studies of different traits should allow
  researchers to disentangle the genetics of potentially related
  traits within individually associated regions.  Formal statistical
  colocalisation testing of individual regions, which requires
  selection of a set of SNPs summarizing the association in a region.
  We show that the SNP selection method greatly affects type 1 error
  rates, with published studies having used methods expected to result
  in substantially inflated type 1 error rates.

  We show that either avoiding variable selection and instead testing
  the most informative principal components or integrating over
  variable selection using Bayesian model averaging can lead to
  correct control of type 1 error rates.  Application to data from
  Graves' disease and Hashimoto's thyroiditis reveals a common genetic
  signature across seven regions shared between the diseases, and
  indicates that in five of six regions associated with Graves'
  disease and not Hashimoto's thyroiditis, this more likely reflects
  genuine absence of association with the latter rather than lack of
  power.  Our examination, by simulation, of the performance of
  colocalisation tests and associated software will foster more
  widespread adoption of formal colocalisation testing.  Given the
  increasing availability of large expression and genetic association
  data sets from disease-relevant tissue and purified cell
  populations, coupled with identification of regulatory sequences by
  projects such as ENCODE, colocalisation analysis has the potential
  to reveal both shared genetic signatures of related traits and
  causal disease genes and tissues.
\end{abstract}


\newpage
\section{Introduction}
\label{sec:introduction}

In recent years, genome-wide association studies (GWAS) have
facilitated a dramatic increase in the number of genetic variants
associated with human disease and other traits such as gene
expression.  Understanding the means by which these variants exert
their effect will aid the design of the detailed functional followup
studies already underway.  Although the causal variants are not
commonly known, multiple traits have been mapped to the same genetic
loci, raising the possibility that the same variants affect multiple
traits either directly or with one trait mediating the other.  For
example, genetic susceptibility to type 2 diabetes across 12 loci
appears mediated by the genetic influence on body mass index
\citep{li11}.  Within individual loci, researchers are examining the
genetic association signals from pairs of traits in parallel, with
similar results interpreted as evidence that the two traits may
colocalise, or share a common causal variant.  These traits may be
eQTL signals across two or more tissues \citep{dimas09,fairfax12},
eQTL and disease signals \citep{nica10,wallace12} or two or more
diseases \citep{2011_cotsapas}.  Distinguishing cases where related
diseases share a common causal variant \emph{versus} those where
neighbouring but distinct variants appear to underlie disease risk in a
region will aid identification of cross-disease and disease-specific
mechanisms.  In addition, comparison of disease and eQTL data has the
potential to reveal both the likely disease causal gene in regions
where 
a number of candidate causal genes exist, and the relevant tissue type
where tissue-specific eQTLs exist.  However, dependence between
genotypes at neighbouring SNPs, caused by LD, means that determination
of colocalisation is not obvious, as there may exist distinct but
neighbouring causal variants for each trait which are mutually
associated.

When these traits are measured in the same individuals, it is possible
to use conditioning to determine whether one trait mediates the other
\citep{li11}.  For example, if both body mass index (BMI) and type 2
diabetes (T2D) have been linked to a SNP, then when including BMI and
the SNP as explanatory variables for T2D, BMI but not the SNP should
show association if BMI is a mediator.  However, when the traits are
measured in distinct samples, or when two traits may share a common
causal variant without one mediating the other, most researchers have
approached the task of looking for colocalisation either by examining
by eye the association signals across a set of common SNPs in the two
data sets \citep{dubois10} or by testing for evidence of residual
association in their available dataset conditional on the most
associated SNP in the other \citep{nica10}.  When full data for both
traits are available, colocalisation may be tested by examining whether
coefficients from regressions of each trait against two or more SNPs
are proportional, as they should be if those SNPs jointly tag a common
causal variant \citep{wallace12,plagnol09}.

We show here that na\"ive application of both conditional and proportional
colocalisation tests may result in substantially inflated type 1
errors, and explore reasons for that inflation.  The inflation cannot
be easily resolved for conditional tests, but we demonstrate two
alternative approaches for proportional testing which result in unbiased
inference.  Finally, we apply these methods to colocalisation testing
of 13 regions shown in Supplementary Table
\ref{tab:power-detect-assoc} which have been associated with one or
both of the autoimmune thyroid diseases, Graves' disease (GD) and
Hashimoto's thyroiditis (HT), using previously published dense
genotyping data \citep{cooper12}.

\section{Methods}
\label{sec:methods}

\subsection{Approaches to colocalisation testing}

We begin by introducing some notation and setting out the details of
the existing approaches to colocalisation testing that are explored in
this paper.  Assume two traits, $Y$ and $Y'$, have been measured in
distinct samples and evidence exists for association of both traits to some genetic region.  Let the region be covered by $p$
SNPs genotyped in both samples, with the genotype matrices denoted by
$X = \left(X_{1},\ldots,X_{p}\right)$ and $X' =
\left(X'_{1},\ldots,X'_{p}\right)$ respectively.  Conditional
approaches begin with identifying the most strongly associated SNPs
for $Y$ and $Y'$, SNPs $k$ and $k'$, say, then examine whether there
is any evidence for association between $Y$ and SNP $k$ conditional on
SNP $k'$.  The null hypothesis is therefore

\begin{equation}
H_0^{\text{cond}} : \quad Y \indep X_{k} | X_{k'}.\label{eq:1}
\end{equation}

Concerned that LD would make interpretation of the conditional test
difficult, \cite{nica10} extended the conditional method
as follows.  For every SNP $j$ generate residuals $R_j$ from a
regression of $Y$ against $X_j$ and test the correlation of $R_j$
and $X_k$ using Spearman's rank correlation test, generating $p$
values $P_j$.  The evidence against the null hypothesis \eqref{eq:1}
is then measured by the rank of $P_{k'}$ in the empirical
distribution, $[P_j]$, generated.  This effectively compares the $p$
value at the test SNP $k$ conditional on SNP $k'$ to that conditioning
on all other SNPs in the region.  However, note that because this
method summarizes evidence for colocalisation by a rank only, there is
no statistical inference attached.  Thresholds for interpreting ranks
 would be expected to depend on
SNP density and LD patterns.

The proportional approach frames the null hypothesis differently.  A set of $q$
SNPs are chosen which are deemed somehow to jointly be good predictors
of one or both traits.  Regressing $Y$ and $Y'$ against these columns of
$X$ and $X'$ respectively produces estimates, $\bb_1$ and $\bb_2$, of
regression coefficients $\bbeta_1$ and $\bbeta_2$, with
variance-covariance matrices $\bV_1$ and $\bV_2$ respectively.  Since
sample sizes are large, the combined likelihood may be closely
approximated by a Gaussian likelihood for $(\bb_1, \bb_2)$, assuming
$\bV_1, \bV_2$ are known and that $\Cov(\bb_1,\bb_2)=0$.
Assuming equal LD in the two cohorts, i.e.\ that the correlation
structure between the SNPs does not differ, \cite{plagnol09} show that
the regression coefficients should be proportional and proposed
testing for a shared causal variant by testing the null hypothesis
$$H_0^{\text{prop}}: \bbeta_1 \propto \bbeta_2,$$
i.e.\ $\bbeta_1=\frac{1}{\eta}\bbeta_2=\bbeta$.   The
chi-squared statistic
\begin{equation}
T(\eta)^2=\bu^T\bV^{-1}\bu \sim \chi^2
\label{eq:x2}
\end{equation}
is derived from Fieller's theorem \citep{fieller54}, where $\bu =
\left(\bb_1-\frac{1}{\eta}\bb_2\right)$ and
$\bV=\bV_1+\frac{1}{\eta^2}\bV_2$. If $\eta$ were known, $T(\eta)^2$
would have a $\chi^2$ distribution on $q$ degrees of freedom.  Plagnol
et.\ al take a profile likelihood approach and replace $\eta$ by its
maximum likelihood estimate, $\hat\eta$, which also minimises
$T(\eta)^2$.  Asymptotic likelihood theory suggests that
$T(\hat\eta)^2$ has a $\chi^2$ distribution on $q-1$ degrees of
freedom.  Alternatively, \cite{wallace12} take a Bayesian approach.
They begin by reparametrising the likelihood in terms of $\theta =
\tan^{-1}(\eta)$ and rewriting the null hypothesis as
$$H_0^{\text{prop}}: \quad \bbeta_1 = \bbeta \cos(\theta); \quad \bbeta_2 = \bbeta
\sin(\theta).$$  This allows calculation of the posterior distribution
of $\theta$, $\mathcal{P}(\theta|\bb_1, \bb_2)$,  assuming
uninformative priors for $\theta$ and $\bbeta$.  Inference is based on posterior predictive $p$ values
\begin{equation}
  \label{eq:ppp}
  \int \! T^*\!(\theta) \mathcal{P}(\theta|\bb_1, \bb_2) \, \mathrm{d}\theta  
\end{equation}
where $T^*\!(\theta)$ is the p value associated with
$T(\tan(\theta))$.  Full mathematical details are given in the
supplementary material of \cite{wallace12}, but it is worth revisiting
here the justification of a flat prior for $\theta$.  If $\beta_1,
\beta_2$ were univariate with Gaussian priors and mean 0, then
$\tan(\theta) = \frac{\beta_1}{\beta_2}$ would have a
$\operatorname{Cauchy}(0,k)$ prior where $k$ is the ratio of the prior
variances of $\beta_1$ and $\beta_2$.  Thus it seems an appropriate
form to consider for the prior for $\theta$ in the multivariate
case. $k$ is unknown, but \cite{wallace12} found that varying $k$
had a negligible effect on the posterior predictive p value
for the sample sizes common in GWAS and eQTL studies (100s to 1000s of
subjects) and we have set $k=1$, implying a uniform prior for
$\theta$, for all analyses in this paper.

Posterior predictive $p$ values have a somewhat different
interpretation than and appear conservative in comparison to standard
$p$ values \citep{rubin84,meng94}.  However, they avoid assuming the
log-likelihood for $\eta$ is approximately quadratic near its maximum
which is not always the case.  In practice,
\cite{wallace12} found standard and posterior predictive $p$ values to
be almost identical in large samples.

$H_0^{\text{prop}}$ is not the same as $H_0^{\text{cond}}$ as it does
not explicitly condition on the most associated SNPs, but is a general
property expected to be true of any pair of traits which share a
common causal variant.  While a shared causal variant should imply
$H_0^{\text{prop}}$ is true at any pair of SNPs, and that
$H_0^{\text{cond}}$ is true if $k=k'$ is the causal variant, the
reverse is not the case as it is possible that two traits have
distinct causal variants in complete LD.  Thus, failure to reject the
null hypothesis indicates only that the data are consistent with a
shared causal variant.

Note that colocalisation testing may be applied equally to case control
data (using logistic regression), expression data (using linear
regression) or to compare case control results against expression
results for a specific gene.  Most commonly, this last approach might
be applied in turn to all genes with a known eQTL signal in the
neighbourhood of the disease association signal.  However, it is
assumed that $\operatorname{cov}(\bb_1,\bb_2)=1$, meaning that case control studies
may only be compared if they do not share a common control group.

\subsection{Choice of SNPs for proportional colocalisation testing}

The choice of SNPs for colocalisation testing will be shown in this
paper to have a considerable influence on the type 1 error rate of
colocalisation tests.  For the proportional approach, two strategies have been applied.  Either
colocalisation has been tested using the pair of SNPs $k$ and $k'$
defined above \citep{plagnol09} or a lasso approach, where SNPs are
first selected in a lasso for one trait, and then additional SNPs are
selected in a further lasso for the other trait \citep{wallace12}. 
However, as shown by Miller \citep{miller84}, any variable selection
method must induce bias in the estimated coefficients ($\bb_1$,
$\bb_2$) if the estimation occurs in the same dataset as the
selection.

The aim of selecting the most informative subsets of SNPs for
proportional colocalisation testing is to minimise the degrees of
freedom of the test, and hence maximize power.  However, unless
independent data are available for variable selection, this increase
in power comes at a cost to type 1 error rate control as shown above.
In this paper, we propose two methods for avoiding this problem.

\begin{figure}
  \centering
  \includegraphics{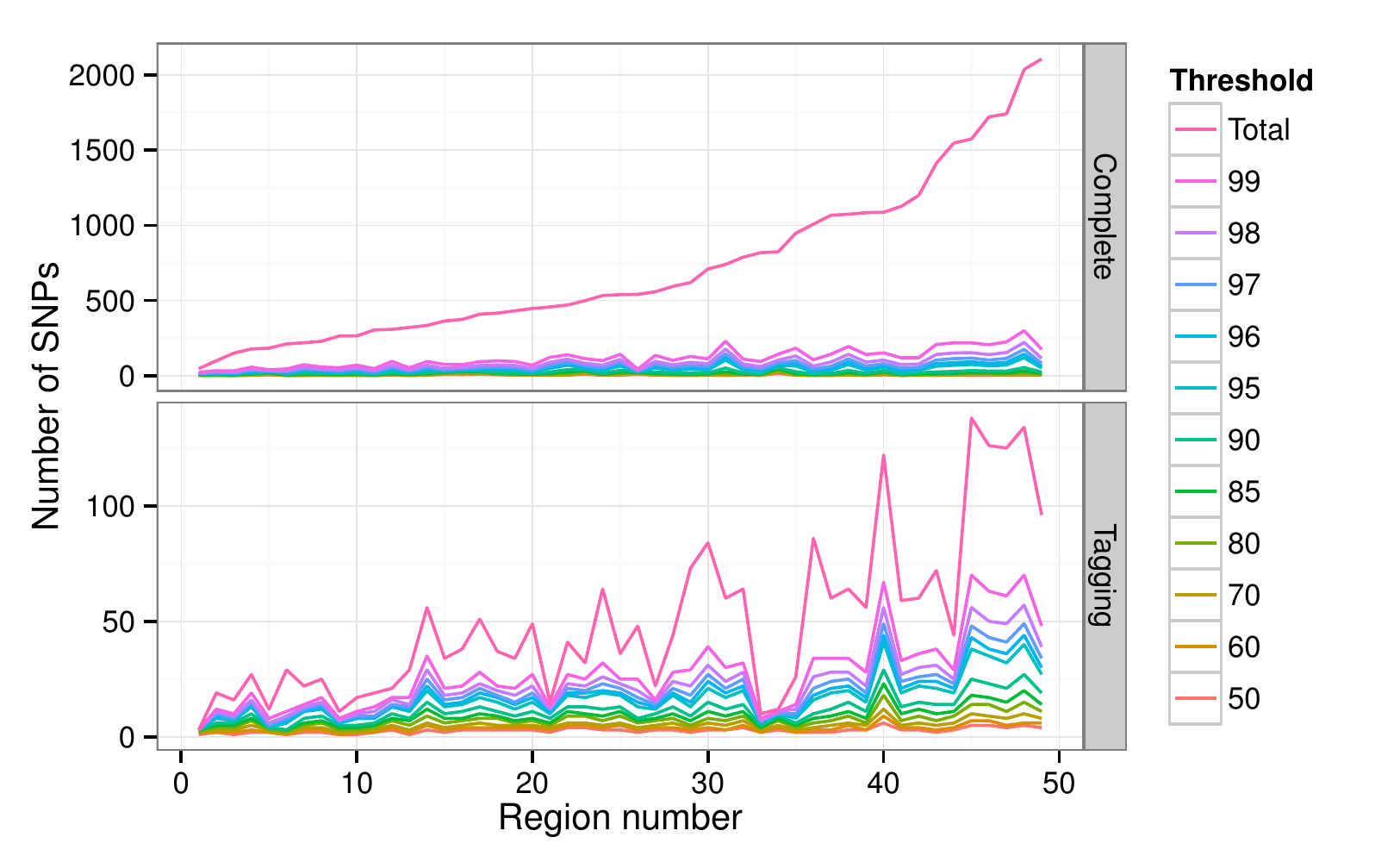}
  \caption[The number of principal components required to capture a
  predefined proportion of the variance]{\textbf{The number of
      principal components required to capture a predefined proportion
      of the variance.}  The 49 regions used for simulation are
    displayed, unlabelled and ordered by the total number of SNPs. The
    majority of variation can be captured by a relatively modest
    number of components even for regions containing large numbers of
    SNPs.  Threshold specifies the minimum proportion of variance
    captured, or ``Total'' for the total number of SNPs in a region.
  }
\label{fig:pc-distribution}
\end{figure}

\paragraph{Summary of genetic variation by principal components}
If the region of interest displays strong LD, a modest number of
principal components (PCs) are generally required to capture most of
the SNP variation (Supplementary Figure~\ref{fig:pc-distribution}) and
we can use a subset of the
most informative components for colocalisation analysis.  Because PCs are by definition
uncorrelated, and because the selection is not based on their
relationship to the traits of interest, the estimated coefficients at
any such subset are unbiased.  To allow PC analysis of two datasets,
we first form a combined genotype matrix, center and scale each SNP,
and then define the principal components.  Colocalisation testing is
performed using the projection of the data onto the transformed basis
for the most important components.  The optimal choice of threshold
defining the ``most important'' components is not obvious, and we
explore that in our simulations.

\paragraph{Bayesian Model Averaging}
Alternatively, we may combine the ideas of Bayesian model
averaging (BMA) \citep{viallefont01} and posterior predictive $p$
values, to treat the model describing the joint association itself as
a nuisance parameter, and average the $p$ values
not just over the posterior for $\eta$, but also over the posterior
for all SNP selection models.  Analogous to equation~\eqref{eq:ppp}, posterior
predictive $p$ values are therefore defined by
\begin{equation}
\text{ppp} = \sum_{m\in \mathcal{M}} p^*\!(m) \mathcal{P}(m)\label{eq:bma}
\end{equation}
where $\mathcal{M}$ is the set of models under consideration, $p^*\!(m)$
is the colocalisation testing $p$ or ppp value under the SNP model $m$,
$\mathcal{P}(m)$ is the posterior probability of model $m$ given the
data and under the assumption that one of ${\cal M}$ is the true
model.  To minimize the degrees of freedom of
the test, we explore all two SNP models and, in the absence of any
independent evidence to favour one SNP over another, we assume the
prior is evenly spread over the set of models.
Approximating the posterior probabilities
by means of the Bayesian Information Criterion approximation
\citep{schwarz78,hoeting99} and discarding highly improbable models at
the outset, this could be done without excessive computational burden
(see Supplementary Material for full details).  

Both the PC and BMA approaches are available in our R \citep{R}
package, coloc, available from the Comprehensive R Archive Network
(\url{http://cran.r-project.org/web/packages/coloc}).

\subsection{Simulation}

We used simulation to demonstrate the effects of variable selection on
the power and type 1 error rate for colocalisation testing.  Full
details are given in supplementary material.  Briefly, we sampled,
with replacement, haplotypes of SNPs with a minor allele frequency of
at least 5\% found in phased 1000 Genomes Project data
\citep{1000genomes12} across all 49 genomic regions outside the major
histocompatibility complex (MHC) which have been identified as type 1
diabetes (T1D) susceptibility loci to date, as summarized in T1DBase
\citep{burren11}.  These represent a range of region sizes and genomic
topography typical of GWAS hits.  We excluded the MHC region which is
known to have high variation, strong LD and exhibits huge genetic
influence on autoimmune disease risk involving multiple loci and hence
requires individual treatment in any GWAS \citep{nejentsev07}.

Using a single ``causal variant'' SNP chosen at random, we sampled
case and control haplotypes according a multiplicative disease
susceptibility model with relative risks of ranging from 1.1 to 1.3 to
represent GWAS data.  To simulate a quantitative trait, and to extend
our exploration to two causal variants in each trait, we selected one
or two ``causal variants'' at random, and simulated a Gaussian
distributed quantitative trait for which each causal variant SNP
explains a specified proportion of the variance.  To reflect our
expectation that this test will be applied in cases in which some
nominal association to a region has already been established, we
discarded datasets in which all single SNP association $p>10^{-4}$.
We either used all SNPs or the subset of SNPs which appear on the
Illumina HumanOmniExpress genotyping array to conduct colocalisation
testing to reflect the scenarios of very dense targeted genotyping
\emph{versus} a less dense GWAS chip.  All analyses were conducted in
R \citep{R} using the \texttt{coloc} package for proportional
colocalisation testing.

\subsection{Colocalisation testing for autoimmune thyroid disease}

An association study of the autoimmune thyroid diseases GD and HT has
recently been completed using the Immunochip for genotyping, which
provides dense coverage of regions of the genome previously associated
with autoimmune disease \citep{cooper12}.  The paper presented a total
of 2285 Graves' disease cases, 462 Hashimoto's disease cases and 9364
controls.  We split the controls randomly into two groups of size
4682, and analysed each of the 13 regions reported to be associated
with one or both diseases \citep{cooper12}.  Missing data were rare,
but regression models require complete genotyping data.  We therefore
imputed missing genotypes by means of multiple regression, as
implemented in the R package snpStats \citep{clayton07_snpmatrix}.  We
conducted proportional colocalisation analysis using the the two
alternative methods set out above.  For the PCs approach, we used
components that captured at least 90\% of the observed genetic
variation.  For the BMA approach, we averaged either over the universe
of all possible two SNP models or that of all three SNP models.

\section{Results}
\subsection{Naive application of colocalisation tests leads to biased inference}

\begin{figure}[t]
  \centering
\includegraphics{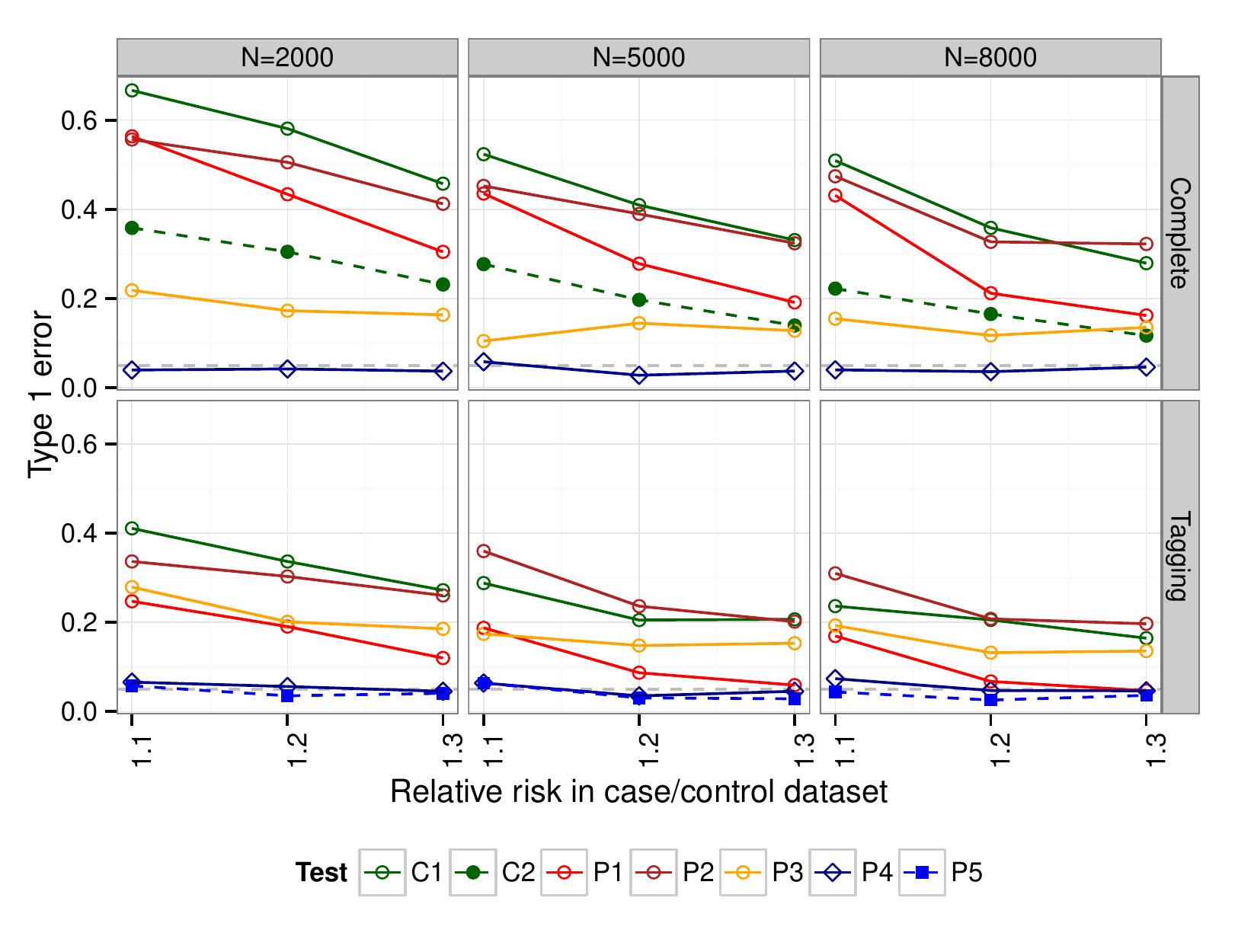}
  \caption{Type 1 error rates in naive colocalisation testing.  A
    nominal type 1 error rate of 5\%, shown by a dashed line, is
    consistently exceeded using conditional colocalisation testing
    conditioning on either the most associated SNP for the other trait
    (C1) or the common causal SNP which is only possible in simulated
    complete genotyping data (C2).  Proportional colocalisation
    testing tends to exhibit lower type 1 error rates, but the excess
    can still be substantial when using the most strongly associated
    SNPs in each dataset (P1); the union of lasso variable selection
    in each dataset (P2) or a two stage lasso variable selection (P3)
    as previously described \citep{wallace12}.  In contrast , type 1
    error rates are well controlled for proportional testing using
    principle components which capture 85\% of the genetic variation
    (P4) or within a Bayesian Model Averaging approach to variable
    selection (P5), even appearing conservative for small effect
    sizes.  Note that Bayesian Model Averaging was not examined in the
    complete genotyping scenario due to computational burden.  The X
    axis shows the relative risk of disease (RR) with columns divided
    according to the number of cases and controls in a case-control
    dataset. Type 1 error rates were calculated by comparing two
    case-control datasets of equal sample and effect size, simulated
    to share a common causal variant.}
\label{fig:type1-naive}
\end{figure}

\begin{figure}[h]
  \centering
  \includegraphics{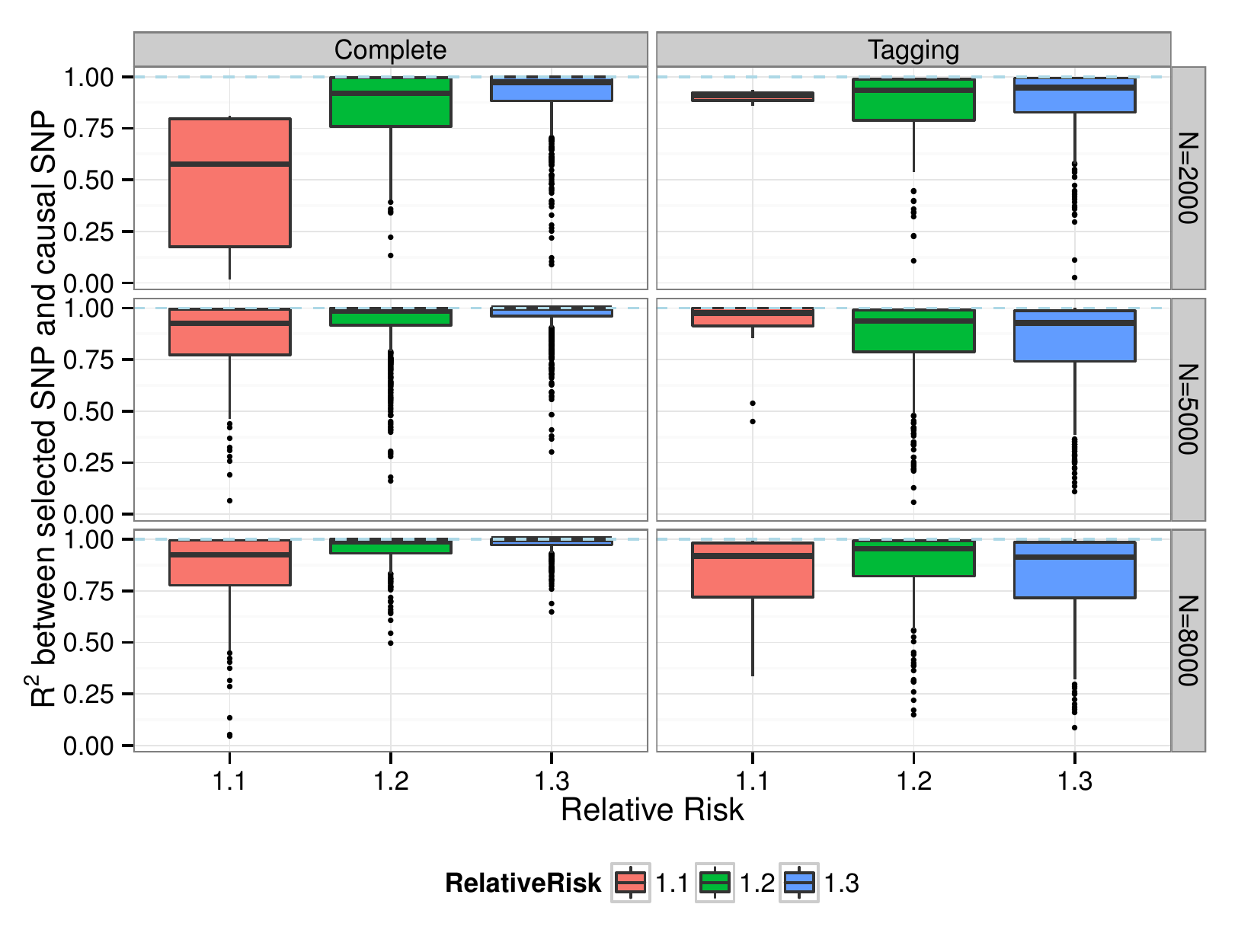}
  \caption[The most associated SNP in a region is not
      necessarily the causal SNP]{\textbf{The most associated SNP in a region is not
      necessarily the causal SNP.}  Boxplots show the distribution of
    $r^2$ between the SNP with the smallest $p$ value (conditional on
    $p<1\times10^{-8}$) and the causal SNP from simulated data, either
    under tagging or complete genotype coverage.  Increasing the
    effect size increases the range of tagging SNPs detectable, and
    hence can have the apparently counter-intuitive result of decreasing
    the correlation between selected and causal SNPs.  However, if
    complete genotype coverage is available, the LD between selected
    and causal SNPs tends to increase with effect size or sample size.}
\label{fig:bias-snps}
\end{figure}

\begin{figure}[p]
  \centering
  \includegraphics{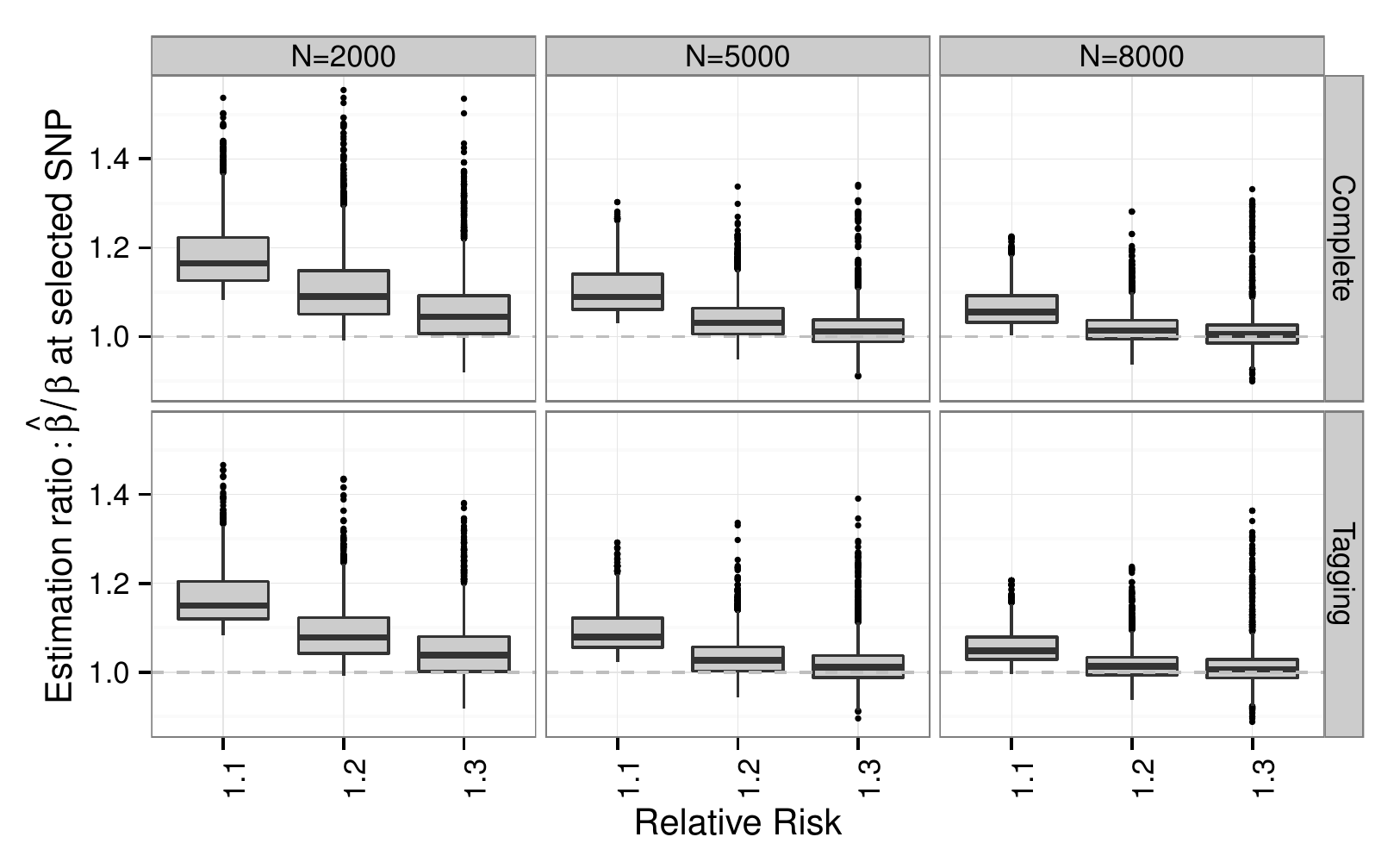}
  \caption{Effect sizes at selected SNPs tend to be overestimated.
    Boxplots show the distribution of the ratio of the estimated
    effect size ($\hat\beta$) to the true effect size ($\beta$) at the
    most associated SNP in a region.  Simulations were conducted for
    samples of N cases and N controls, with a relative risk at a
    randomly selected ``causal SNP'' of 1.1, 1.2 or 1.3, under either
    a complete genotyping scenario (all SNPs in 1000 Genomes, top row)
    or the subset of SNPs appearing on the Illumina Human Omni Express
    chip chip (``Tagging'', bottom row).  Estimated effects are more
    likely to be biased for smaller effect sizes and sample sizes.}
\label{fig:bias-effect}
\end{figure}

The choice of SNPs to use for testing can induce  bias for two
reasons.  First, selecting the ``most associated'' SNP on the basis that the
evidence for its association is strongest amongst all SNPs tested does not
guarantee either that it is the causal SNP or even the best proxy.  %
Random variation and LD mean that evidence
for association may peak at an alternative SNP even when the causal
SNP is included in the genotyping panel, a bias which is more
pronounced for weaker effects and smaller sample sizes (Supplementary Figure
\ref{fig:bias-snps}).  Second, although it is well known that
regression coefficients are unbiased estimates of population effects,
this property does not hold after variable selection \citep{miller84}, an
effect which has been referred to as ``Winner's curse'' in genetics
\citep{goering01,lohmueller03}.  Choosing SNPs on the basis of their significance or some other
measure of strength of association induces a bias away from the null -
i.e.\ coefficients of selected SNPs are expected to overestimate the true
effect - and again the effect is more pronounced for smaller sample
sizes and weaker effects (Supplementary Figure
\ref{fig:bias-effect}).  %
These two biases mean that, in conditional testing, there is likely to
be some residual association between the phenotype $Y$ and the
remaining genetic markers after conditioning on the selected SNP $k'$
because (1) the conditioning SNP $k'$ may not capture all the true association
and (2) the estimated effect at the tested SNP $k$ tends to be an overestimate.
The result is very poor control of the type 1 error rate
(Figure~\ref{fig:type1-naive}, track C1).  Conditioning on the common
causal variant rather than the most associated SNP (which is only
possible in simulation studies) reduces the bias by removing the
SNP selection problem, but does not eliminate it due to the
overestimation of effect size (Figure~\ref{fig:type1-naive}, track
C2).

As seen in Supplementary Figure~\ref{fig:nica}, Nica's score tends
towards to 1 for traits that share a causal variant and is uniformly
distributed on $[0,1]$ for distinct unlinked causal variants. Its
distribution is increasingly skewed towards 1 as the LD between
distinct causal variants increases.  This makes sense if one considers
that the case of two distinct variants in some LD lies partway between
the extreme cases of distinct linked causal variants and a single
common causal variant, which is equivalent to distinct causal variants
in complete LD.  The effect of using the most associated SNPs for
testing compared with using the true causal SNPs is to reduce the skew
towards higher rank scores as the $r^2$ between variants increases.
Thus, Nica \emph{et.\ al}'s extension \citep{nica10} is useful if
searching for most likely colocalisation signals within a set, but as
it avoids formally testing a null hypothesis, and because the scale
against which to interpret the rank score is likely varies according
to effect size, it does not provide a means to assess evidence for or
against colocalisation at a given locus of interest.

We show here that neither published method of SNP selection in
proportional testing maintains control of the type 1 error rate
(Figure \ref{fig:type1-naive}, tracks P1, P2 and P3), although the
bias is less extreme than for conditional testing.  The two-step lasso
selection defined above does reduce bias compared to independent lasso
selection in the two datasets, but, perhaps counter-intuitively, leads
to greater bias than simply testing the pair of most associated SNPs
$(k,k')$ when only tagging genotypes are available and effect sizes
are large (relative risk $\sim$ 1.3).  This is because, in this
situation, lasso may select SNPs which are apparently weakly
associated (either truly or through random noise) at which, as
demonstrated in Supplementary Figure \ref{fig:bias-effect}, effect
estimates are more strongly biased.

\subsection{Proper control of type 1 error rates}
\label{sec:altern-meth-snp}

\begin{figure}
  \centering
  \includegraphics{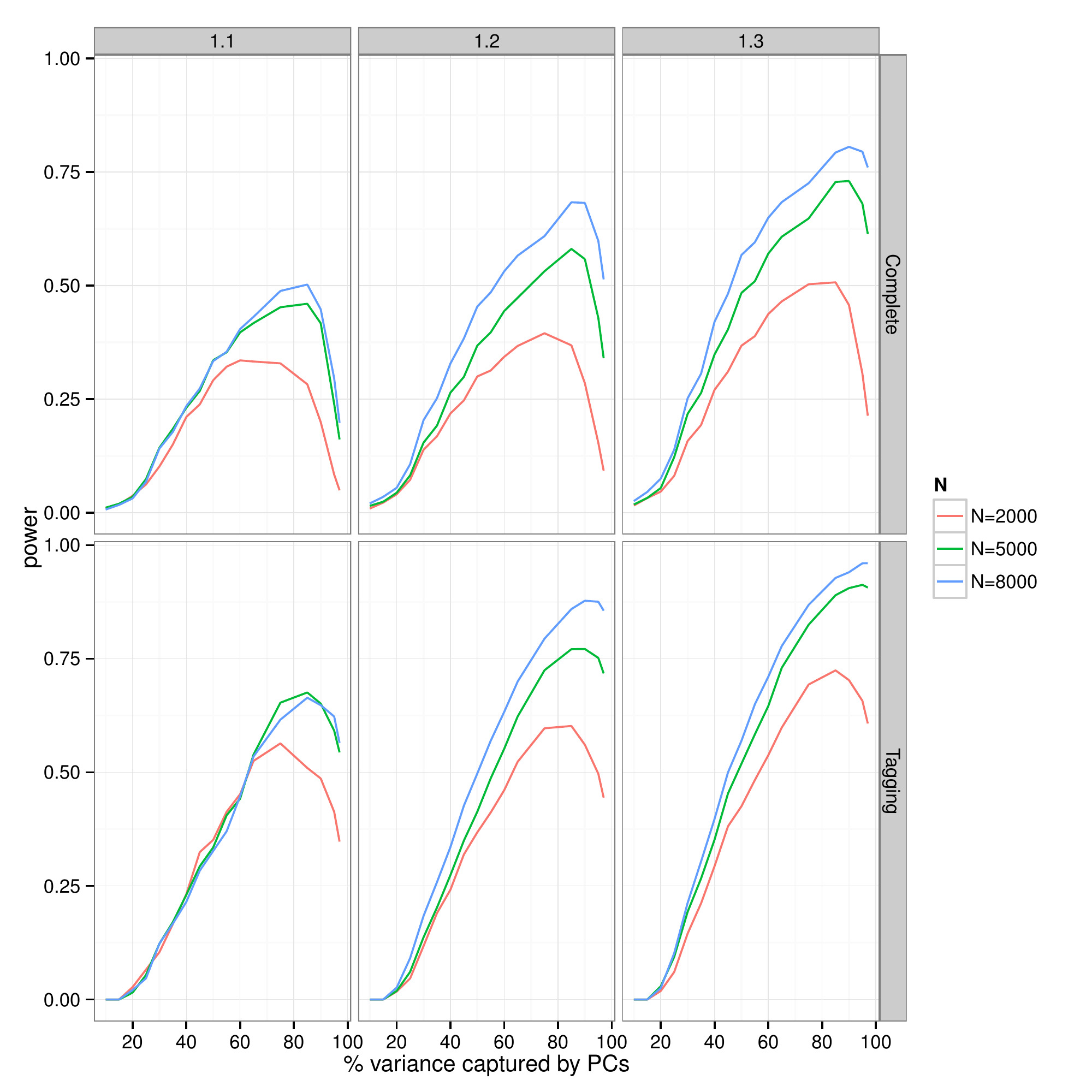}
  \caption[Power using colocalisation testing of principal components
  according to the proportion of genotype variance
  captured]{\textbf{Power using colocalisation testing of principal components
  according to the proportion of genotype variance
  captured.}  Power is shown for all simulated datasets where the
$r^2$ between the causal SNPs was less than 0.5.}
\label{fig:pc-power}
\end{figure}

\paragraph{Principal components}
When using principal components to summarize the genetic variation in
a region, it is not obvious how many components are required.  As more
components are selected, more information about the genetic variation
in a region is captured, and hence we are more likely to accurately
capture the signal of any causal variants.  However, successive
components add decreasing amounts of information whilst still adding
another degree of freedom.  At some point the negative effect of
increasing degrees of freedom will outweigh the positive effect of
increasing information, and we were concerned that the optimal test
may depend heavily on the threshold used to determine the number of
components selected.  Instead, power seemed broadly acceptable once
components capturing 70-90\% of variation were selected (Supplementary
Figure~\ref{fig:pc-power}).  In our 49 test regions, 70\% of the
variance could be captured by selecting an average of 7 (range 2-18)
or 9 (range 3-44) components under a tagging or complete genotyping
approach. 

\begin{figure}[t]
  \centering
  \includegraphics[width=\textwidth]{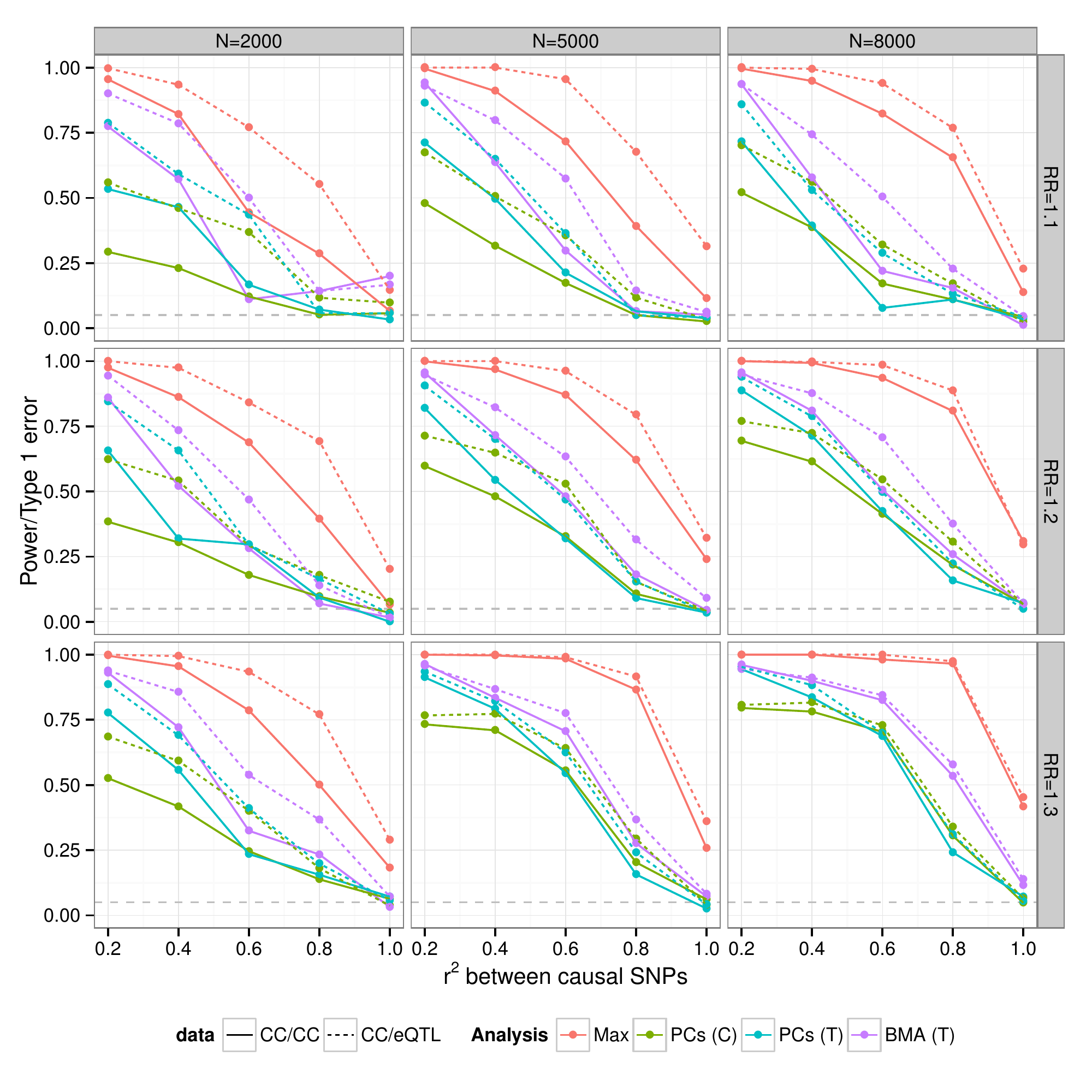}
  \caption{Power for proportional colocalisation analysis using PC
      or BMA approaches.  The theoretical maximum power (Max) is
    calculated by proportional colocalisation testing using the two causal
    variants which are known in simulated data and show that the
    predominant determinant of power is the $r^2$ between the
    variants, with power decreasing as LD increases.  When the causal
    variants are not known, power decreases under either a PC or BMA
    approach.  The X axis shows the maximum $r^2$ between the causal
    variants, i.e.\ $r^2$ has been categorised into 5 groups: $[0,0.2]$,
    $(0.2,0.4]$, $(0.4,0.6]$, $(0.6,0.8]$, $(0.8,1.0]$. N is the
    number of cases and controls in a case-control dataset with
    relative risk of disease RR.  Power is shown for comparing two
    case-control studies with equal sample numbers and effect sizes
    (solid lines) or for comparing a case-control study to an eQTL
    study of 1000 samples where the causal variant explains 30\% of
    the variance of the expression.  The PC approach was implemented
    by selecting the smallest subset of components which captured 85\%
    of the genetic variance.  We considered only tagging genotype
    scenarios to reduce computation time.}
  \label{fig:power}
\end{figure}

We found type 1 error rates were controlled across the range of
thresholds explored, but show for illustrative purposes the results
when we fixed the threshold at $\geq$90\% of genetic variation  
(Figure~\ref{fig:type1-naive}). 
We examined power to detect departure
from colocalisation using simulations in which the causal variants are
distinct for two traits but placed no restrictions on the LD between
these variants.  We first examine the theoretical maximum power of the
test by testing the two causal variants themselves, which are known in
a simulation study but not in real data (Figure~\ref{fig:power}).  As
might be expected, power increases with sample size and effect size,
but is negatively correlated with the $r^2$ between the causal
variants, and is maximum when the two are completely unlinked
($r^2=0$).  When using PCs, the power is reduced reflecting the loss
of information in not knowing these causal variants, but the loss is
greatest for complete genotyping scenarios, suggesting that we may
be selecting too many components in the case of complete genotyping
and emphasizing the difficulty in choosing one optimal threshold for
all studies and regions.

\paragraph{Bayesian Model averaging} 
Because of its computational burden for simulations, we only consider
the BMA approach under a tagging genotyping scenario.  This
demonstrates good control of the type 1 error, even tending to be
mildly conservative, as has previously been reported when posterior
predictive $p$ values are interpreted similarly to standard $p$ values
\citep{meng94}.  Despite the slightly more conservative type 1 error
rates, the BMA approach appears more powerful than the PCs approach
(Figure~\ref{fig:power}), which presumably reflects the greater
degrees of freedom required for the PCs approach.

\subsection{Sensitivity to the assumption of equal linkage disequilibrium}
\label{sec:effect-depart-from}

\begin{figure}[h]
  \centering
\includegraphics[width=1.1\textwidth]{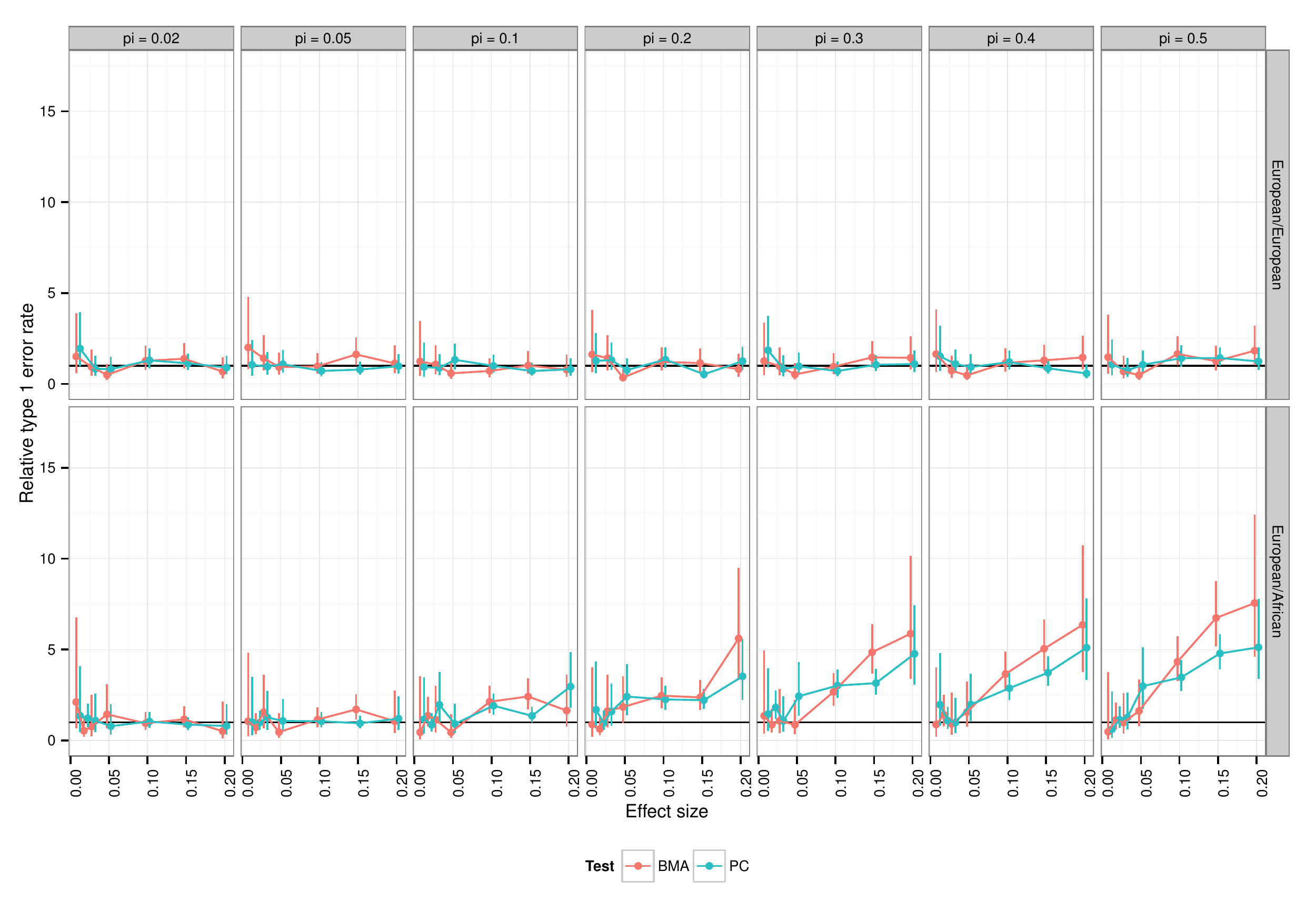}
\caption{Departure from assumption of equal LD structure.  Each plot
  reflects simulations in which a single, common causal variant
  explains a fixed proportion of variance of two quantitative traits,
  shown on the x axis, each available in a sample of 1,000
  individuals.  In the top row, all haplotypes in the first dataset
  and $(1-\pi)$ of the haplotypes in the second dataset were sampled
  from the European CEU, GBR and FIN populations, and the remaining
  $\pi$ in the second dataset from the alternate European TSI and IBS
  populations.  In the bottom row, we used the same strategy but
  sampling either from all European populations (CEU, GBR, FIN, TSI
  and IBS) or from a mixture of these European populations and the
  African ASW , LWK and YRI populations.  The y axis shows the
  relative type 1 error rate - the ratio of the estimated rate for the
  given scenario and the estimated rate for the equivalent scenario
  with no mixing.  Because these are ratios, there is rather less
  certainty than for other plots and 95\% confidence intervals
  calculated by means of the delta method are shown for each point.
  Analysis was conducted by proportional testing using either a PC
  approach with number of principal components selected to capture
  90\% of genetic variation or a BMA approach, averaging over the
  space of either all possible two SNP models.  We considered only
  tagging genotype scenarios.  CEU=Utah Residents (CEPH) with Northern
  and Western European ancestry; GBR=British in England and Scotland;
  FIN=Finnish in Finland; TSI=Toscani in Italia; IBS=Iberian
  population in Spain; ASW=Americans of African Ancestry in SW USA;
  LWK=Luhya in Webuye, Kenya; YRI=Yoruba in Ibadan, Nigera}
  \label{fig:admix}
\end{figure}

The proportional colocalisation test assumes identical patterns of LD
in the two datasets so that the effect of a shared causal variant is
proportional across any set of SNPs.  To explore its sensitivity to
this assumption, we considered sampling haplotypes for one dataset
from a subset of European populations, and for the other dataset from
a either a mixture of European populations or a mixture of European
and African populations.  As might be expected, for strongly admixed
datasets, the control of type 1 error rate is lost, with type 1 error
rates up to 8 fold that seen under the case of no mixing
(Figure~\ref{fig:admix}).  However, it is perhaps surprising that the
effect of mixing between two European populations, or mixing very
small proportion of African haplotypes ($\sim 5\%$) into a mainly
European population, is barely detectable at the sample sizes of 1,000
used, and indicates that the method is not very sensitive to small
departures from the assumption of equal linkage disequilibrium.  Of
course, as with any genetic analysis, it remains sensible not to rely
on this property, but to formally examine the evidence for population
structure and exclude obviously outlying samples.

\subsection{The case of multiple causal variants}
\label{sec:two-causal-variants}

\begin{figure}[h]
  \centering
\includegraphics{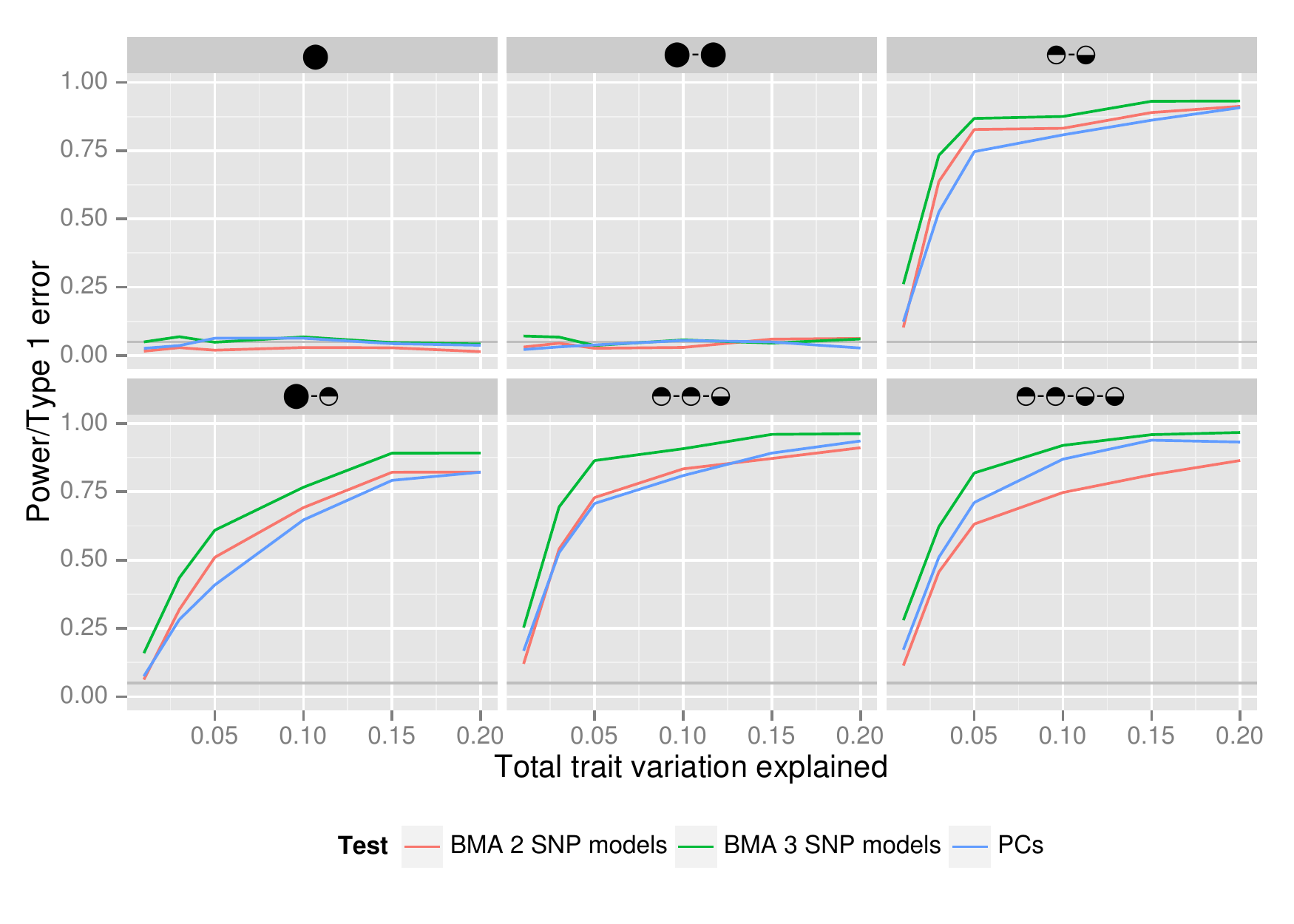}
  \caption{Effect of more than two causal variants.  Each plot
    reflects simulations in which causal variants in total explain a
    fixed proportion of variance of two quantitative traits, shown on
    the x axis, each available in a sample of 1,000 individuals.  The
    total number of causal variants is shown by the number of circles
    above each plot, with full circles indicating a causal variant
    shared by both traits and half shaded a causal variant associated
    with one trait or the other.  Analysis was conducted by
    proportional testing using either a PC approach with number of
    principal components selected to capture 90\% of genetic variation
    or a BMA approach, averaging over the space of either all possible
    two SNP or three SNP models.  We considered only tagging genotype
    scenarios.}
  \label{fig:multi}
\end{figure}

So far, we have only considered the case of a single causal variant
for each trait.  But the proportional test makes no assumption about
the number of causal variants, only that their effects are
proportional.  Figure \ref{fig:multi} shows that in the case of eQTL
data with two shared causal variants, having equal effects on each
trait, type 1 error rates are still controlled.  It has been reported
that genes may exhibit a common cross-tissue eQTL, located proximal to
the gene, as well as distinct tissue specific eQTLs in more distal
locations \citep{brown_integrative_2012}.  We were therefore
interested to explore the case where our two traits share one causal
variant, but one or both are also under the influence of additional,
distinct variants.  Testing a single hypothesis of colocalising
\emph{versus} not colocalising variants cannot capture the complexity
of this situation, but it is instructive to explore the test's
expected behaviour in order to allow proper interpretation with real
data where the number, and sharing configuration of causal variants is
unknown.  Figure~\ref{fig:multi} shows that, under a tagging
genotyping scenario, the proportional test tends to reject the null of
colocalisation in the case of any distinct causal variants, even in
the presence of an additional shared variant, although with slightly
less power than when there is no shared variant.

\subsection{Varying the number of SNPs in Bayesian Model Averaging models}
\label{sec:varying-number-snps}

So far, also, we have assumed that it was enough to consider only the
universe of two SNP models when applying our BMA approach.  The
motivation for this was that a two SNP model leads to a one degree of
freedom test, and might therefore be expected to maximise power.  We
examined the effect of averaging over either all two SNP or all three
SNP models in the context of the above multiple causal variant
simulations.  This shows that type one error rates are similar, that
power is similar for two SNP models, but that power can be increased
by averaging over three SNP models compared to two SNP models,
particularly when there are really three or more distinct causal
variants (Figure~\ref{fig:multi})

\subsection{Application to colocalisation testing of Autoimmune Thyroid Diseases}

Existing evidence suggests that a single locus may contain variants
which predispose to any one of multiple diseases, e.g.\ the
non-synonymous C1858T SNP in \emph{PTPN22} is associated with
rheumatoid arthritis and T1D \citep{stahl10,barrett09}, or distinct
variants which predispose to different diseases, e.g.\ distinct variants
in \emph{IL2RA} are associated with T1D and multiple sclerosis
\citep{maier09:il2ra,martin12}.  We used the proportional colocalisation
approach outlined above to examine the disease signals for the 
autoimmune thyroid diseases HT and GD from a recent dense genotyping
study \citep{cooper12}.  

\begin{figure}
  \centering
  \includegraphics[trim=0cm 45mm 0cm 0mm,clip=TRUE,width=\textwidth]{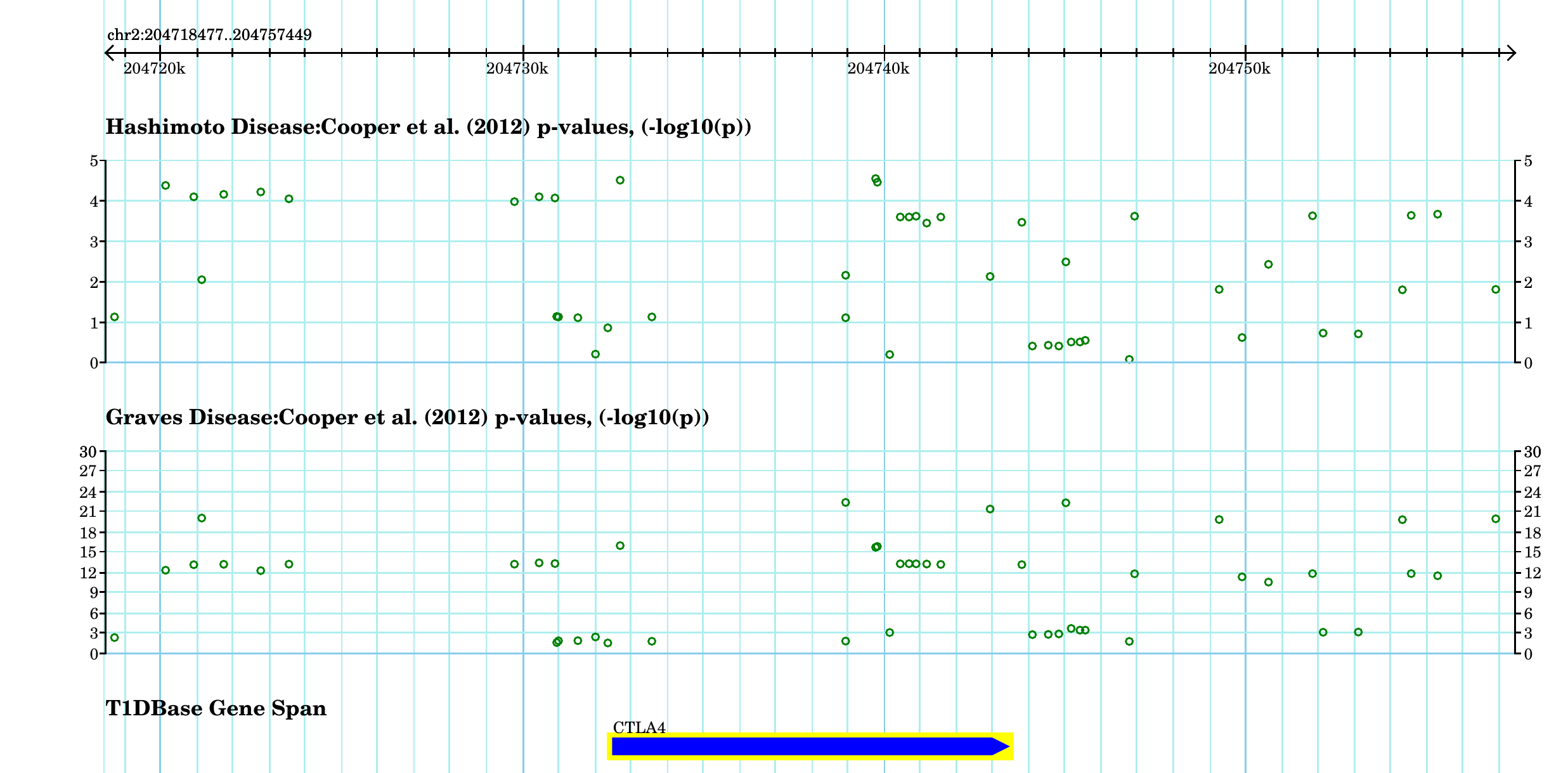}
  \caption{Single SNP p values for GD and HT in the
    \emph{CTLA4}/\emph{ICOS} region on 2q33.2.}
  \label{fig:t1dbase}
\end{figure}

We first examined the seven regions where a significant single SNP
effect has been identified in both diseases, i.e.\ at study-wide
significant levels for GD ($p<1.1\times 10^{-6}$, a permutation
derived threshold specific to this study) and at a nominal
significance threshold of $p<0.05$ for HT.  Six of these display no
evidence against colocalisation (all posterior predictive $p>0.01$),
the exception being 2q33.2/\emph{CTLA4}/\emph{ICOS} in which the ppp
value for the BMA approaches is $6\times10^{-3}$ (two SNP models) or
$8\times10^{-5}$ (three SNP models).  In this region, the profile of
the single SNP $p$ values do differ (Supplementary Figure
\ref{fig:t1dbase}), but it would require larger sample sizes to
confidently conclude that the two diseases have different causal
variants in this region given the number of tests completed.

\begin{figure}[t]
  \centering
\includegraphics[width=\textwidth]{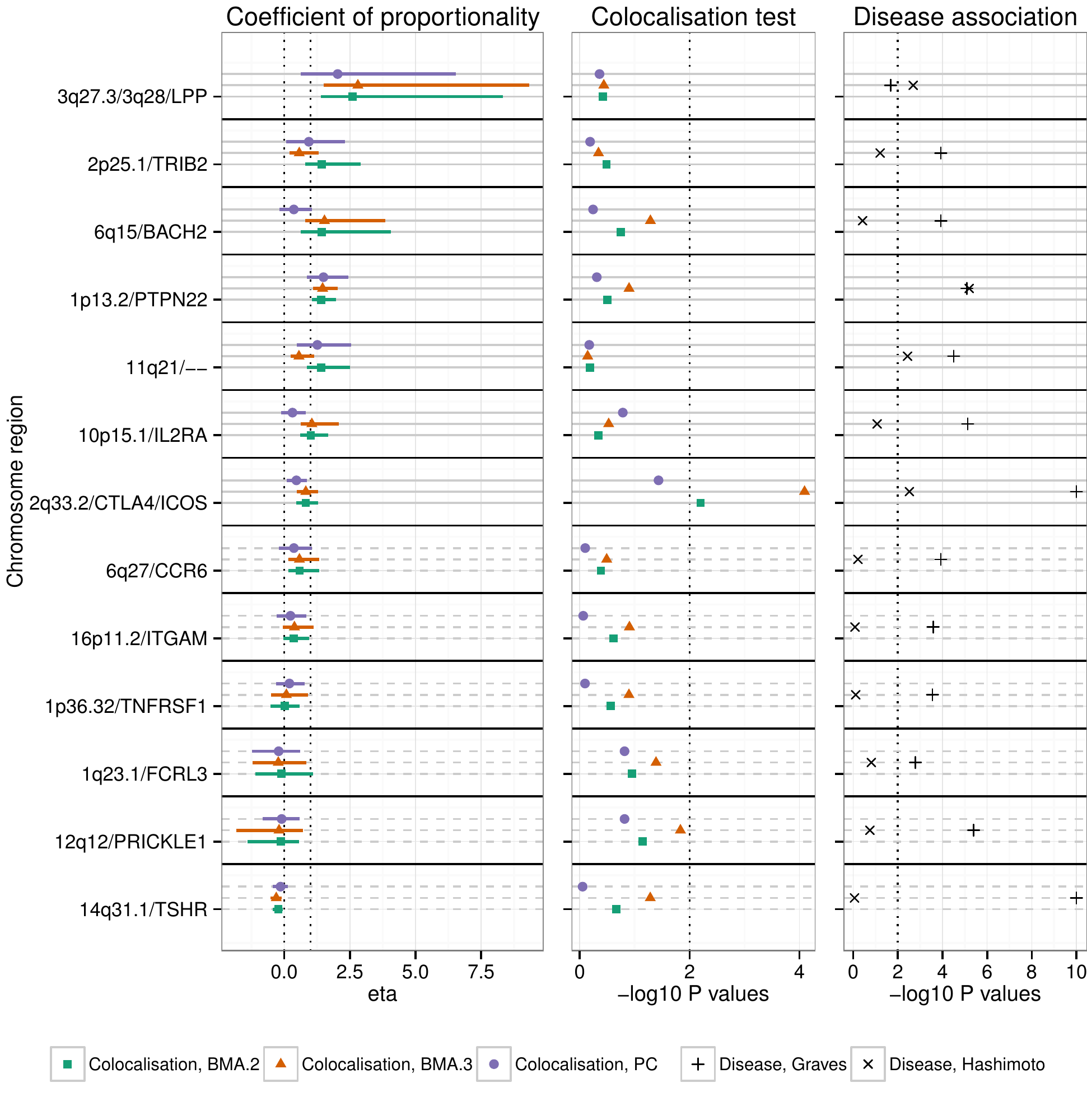}
\caption{Colocalisation analysis of Graves' and Hashimoto's diseases.
  Regions are labelled by chromosome and likely candidate gene(s) and
  arranged so that the top seven regions showed marginally significant
  association with both GD and HT ($p<0.05$) and the bottom six with
  just GD in the published single SNP analysis \citep{cooper12}.  The
  left panel shows the estimate of the coefficient of proportionality,
  $\eta$, for the estimate of the ratio of effect sizes in HT compared
  to GD, and its 95\% credible interval, calculated using either the
  PCs approach (PC), or BMA averaging over two SNP models (BMA.2) or
  three SNP models (BMA.3).  The middle panel shows the the evidence
  against colocalisation using either the $-\log_{10}(p)$ value (PCs)
  or the posterior predictive $-\log_{10}(p)$ (BMA).  The right panel
  summarizes the evidence for association of the region with each
  disease using $-\log_{10}(p)$ values for the association analysis of
  Graves' and Hashimoto's using the selected principal components for
  the PCs approach.  The $-\log_{10}$ scale has been truncated at 10 so
  that more extreme $p$ values are displayed at
  $-\log_{10}(p)=10$. For the PC approach, testing was based on the
  smallest subset of components that captured 90\% of the genetic
  variance.  }
  \label{fig:aitd}
\end{figure}

The coefficient of proportionality, $\eta$, can be usefully
interpreted when analysing two diseases.  Two values of particular
interest are $\eta=0$ which would indicate no effect in HT given an
effect in GD and $\eta=1$ which indicates equal effects in each
disease.  In most of the seven regions, the credible interval for
$\eta$ includes 1 (Figure~\ref{fig:aitd}), the exceptions being 2q33.2 and 10p15.1, where it
ends just below 1 using a PCs approach, and 3q27.3/3q28/\emph{LPP} where it starts
just above 1 on either BMA approach.  

Turning to the the six regions where there is evidence of association
in only GD, 
we do not expect to see any departure from the null of colocalisation,
without evidence of association to both traits, and indeed all
posterior predictive $p>0.01$.  However, our estimate of $\eta$ helps
infer whether this reflects a lack of power or genuine absence of
association for HT.  We evaluated the credible intervals for $\eta$ in
each region and across four of the six regions, the credible interval
for $\eta$ includes 0, whilst in only one does the credible interval
include 1 for all approaches.  In one 14q31.1/\emph{TSHR}, the
credible interval ends just \emph{below} 0 for the BMA approaches,
whilst for 6q27/\emph{CCR6}, it starts just above 0 and includes 1 for
all methods.  \emph{TSHR} represents the primary autoantigen in
hyperthyroidism of GD \citep{brand10}, so is unlikely to be involved
in HT.

\section{Discussion}

There are two sources to the bias in colocalisation testing presented
above.  The problem of variable selection is well studied in
statistics generally \citep{miller84} but has perhaps been neglected in
statistical genetics, where the aim has been to detect convincing
association to a region, rather than pinpoint the causal variant,
particularly as most datasets to date have included an incomplete
selection of variants in any region.  Selecting SNPs which do not
fully capture the trait association will affect conditional
colocalisation testing because some residual association must remain
after conditioning.  On the other hand, it should not bias proportional
testing as the aim there is to test for proportionality of effect size
rather than evidence of residual effect.  This may explain the
substantially higher error rates for naive conditional testing
\emph{versus} naive proportional testing seen in Figure
\ref{fig:type1-naive}.

The bias in effect size estimates affects both methods, however.  In
genetics, we are familiar with ``Winner's curse'', which causes effect
estimates which are examined conditional on the associated $p$ value
being below some significance threshold to be biased away from the
null.  Some attempts to correct this effect size bias have been made,
either by modelling a selection procedure defined as a single SNP
exceeding a predetermined level of significance \citep{zollner07}, or
by bootstrapping which can in theory account for the full selection
strategy \citep{sun11}.  We explored both approaches, but found
neither led to unbiased or even nearly unbiased inference (data not
shown).  For \cite{zollner07}, this failure is presumably down to the
discrepancy between correcting for a p value that exceeds some
threshold and selecting SNPs with the minimum p values.  In the case
of \cite{sun11}, it is possible that single loci do not contain
sufficient information for a bootstrap based correction; the
corrected estimates tended to be biased in the opposite direction,
suggesting the method was over-correcting.  

Our proposed solution is
to use proportional testing and either avoid variable selection
altogether by using the PCs which capture the majority of genetic
variation in the region under test, or integrate over the variable
selection using BMA.  Either method maintains type 1 error, and the
BMA approach appears more powerful than the PC approach, although both have
reduced power compared to the hypothetical scenario of being able to
test the causal variants themselves.  Recall from
equation~(\ref{eq:bma}) that the ppp depends on both the posterior
probabilities of individual SNP models and the degree of departure
from the null under each model, $p^*(m)$.  If the posterior
probability was spread broadly over the model space and $p^*(m)$ varied
considerably across models, the distribution of
ppp would be far from uniform, with too few observations in the tails.
The fact that we observe only mildly conservative type 1 errors,
even in the weakest associations considered here (case control sample
sizes of 2000, and relative risks of 1.1), probably reflects our
requirement in simulating our datasets that some nominal level of
association must be observed for at least one SNP in a region.  This
means that, for our simulated data, the posterior probabilities tend
to be mainly focused on a small subset of models, and these models
tend to contain sets of SNPs related by LD so that $p^*\!(m)$ does not
vary substantially across the set of models on which most of the
posterior probability is concentrated.  Whether caution is needed in
applying this approach in smaller datasets depends somewhat on a
researcher's view.  Certainly, it is unlikely that the ppp
distribution would be uniform if small samples with weak associations
were used, and therefore power to detect departure from colocalisation
would be reduced.  However, the alternative to colocalisation being
tested is not just that of no colocalisation, but implicitly that of
distinct causal variants in two traits.  If there is weak or only
limited evidence for association with either trait, then it does seem
appropriate that one cannot reject a null of colocalising causal
variants in favour of an alternative distinct causal variants.  On the
other hand, if evidence for association has been independently established, a researcher may be sure that there are causal
variants for each trait although there is only weak evidence in the
currently available datasets, in which case they may wish to explore
calibrating the ppp distribution, for example by means of simulation
\citep{hjort06}, so that its null distribution is uniform between $[0,1]$.

Regions are typically defined after GWAS studies according to genetic
distance in order to describe the physical region within which a
causal variant tagged by the association is expected to lie. T1DBase
uses a definition of 0.1cM surrounding any single SNP with $p\times
10^{-8}$ and we used a 0.1cM window around the most associated SNP in
our AITD analysis.  If regions were defined more broadly, one might
expect the BMA method to be relatively unaffected, as SNPs beyond the
boundary of the current regions would not show much evidence for
association.  On the other hand, we expect the power of the PC
approach would decrease, as the number of components, and therefore
the degrees of freedom of the test, would increase without any
compensatory increase in information.

We have proposed alternative approaches to colocalisation testing, but
how should a researcher choose which approach and the parameters
specific to that approach?  Our view is that we would use PCs as a
first pass if many loci are under consideration, to prioritise regions
for more detailed analysis.  The optimal number of components is
unknown, but a number that captures something in the region of
70--90\% of genetic variation seems acceptable.  However, if evidence
for association is strong in both datasets, BMA is a more powerful
approach and there seems little reason beyond computational expense to
prefer averaging over the universe of two SNP models to that of three
SNP models.  However, in large regions, a relatively modest
number of SNPs can cause the number of three SNP models to be
infeasible.  For example, 100 SNPs can generate 4,950 two SNP models
but 161,700 three SNP models.  We would not suggest exploring four SNP
models because that seems likely to spread the posterior probabilities
very thinly, and we haven't explored combining two and three SNP
models because it is unclear what prior weight should be given to a
two SNP compared to a three SNP model.

As an example application of our proposed approaches, we analysed 13
loci associated with the autoimmune thyroid diseases GD and HT.
Reassuringly, inference was broadly similar regardless of method, and
we and showed that in the seven regions where a locus has been
associated with the two diseases, the data are generally consistent
with common causal variants exerting an equal effect on each disease.
In regions previously significantly associated with only GD, posterior
predictive $p$ values are unlikely to detect any departure from
colocalisation for reasons described above, but here the Bayesian
approach does allow us to use make some inference using the posterior
distribution of $\eta$.  Given the relatively smaller number of HT
cases (462) compared to GD (2285), it might be expected that many of
the loci only associated with GD have failed to be reach significance
for HT due to lack of power.  Estimates of power to detect association
with HT under the assumption of equal effects in GD and HT are broadly
similar across the seven regions associated with both diseases and the
six regions associated with GD only (Supplementary Table
\ref{tab:power-detect-assoc}), but these are likely to be over
optimistic due to the expected bias in the GD effect size estimates.
However, for five of these six regions, the evidence suggests that
failure to detect association with HT is more likely to be due to a
discrepancy in effect size, with the effect considerably larger in,
and possibly specific to GD, compared to HT, rather than a lack of
power.  However, while the data suggest that \emph{LPP} may
have a stronger effect on GD compared to HT, and that \emph{CCR6} may
be an undetected HT locus, the small sample size for HT prevents us
from drawing this conclusion with any confidence.

There is a pressing need for more widespread use of formal
colocalisation testing.  Researchers are turning to eQTL data to
interpret GWAS results by simply considering whether an eQTL SNP is
associated with any disease \citep{nicolae10}, visually
\citep{trynka11}, by conditional testing \citep{nica10}, by naive
application of a proportional colocalisation test \citep{plagnol09,wallace12}
or by attempting to integrate disease and gene expression association
signals in networks \citep{hao12}.  Where colocalisation tests have
been naively applied, we expect the null hypothesis of colocalisation
has been rejected too readily, although this will affect loci with
small and moderate effects to a greater degree than those with large
effects.  Thus, for our earlier analysis of colocalisation between T1D
and monocyte gene expression signals \citep{wallace12}, the list of
loci compatible with colocalisation are likely correct, but some loci
were probably erroneously rejected, and
re-analysis of these data will be required.

In the case of network analysis, results can be difficult to reconcile
with simple representations of the data.  For example, integration of
lung expression and asthma genetic association data led to the
identification of \emph{GSDMA} as the most likely causal gene for
asthma in the 17q21 region \citep{hao12}, despite a graphical
representation of the data showing that the SNPs most strongly
associated with \emph{GSDMA} expression were relatively weakly
associated with asthma, and, \emph{vice versa}, that the SNPs most
strongly associated with asthma showed relatively weaker levels of
association with \emph{GSDMA} expression compared to the strongest
signals.  The asthma association on the 17q21 region was one of the
first cases of explicitly using expression data to interpret disease
association, with the association with asthma initially attributed to
\emph{ORMDL3} based on expression data from EBV transformed cell lines
\citep{moffatt07} and subsequently to \emph{GSDMB} from a reanalysis
of the same data \citep{moffatt10}.  Candidate gene hypotheses have
been constructed for all three genes.  The lung expression data have
greater potential for revealing the underlying gene, but, to hold
confidence in results of analyses, particularly when the results
contrast with simple visual inspection of the data, requires careful
examination of the properties of the statistical method used.

Given the tissue-specific nature of many eQTLs identified to date
\citep{dimas09,fairfax12}, there is a need for more large, publicly
available eQTL datasets in a variety of disease relevant tissues and
purified cell subsets to support the interpretation of existing GWAS
data.  Although expression data is typically shared after the
publication of an eQTL study, we note that the genetic data must also
be made available to allow full integration of eQTL and disease
signals at shared loci.  The increasing abundance of substantial GWAS
datasets and the increasing availability of
large eQTL datasets \citep{fairfax12,fu12,hao12}, together with our
reassurance that it is possible to conduct these tests whilst
maintaining type 1 error rates and the availability of software in our
R package will facilitate more widespread formal colocalisation
testing.  Integration of genetic association data has the potential to
refine understanding of underlying genetic mechanisms and aid in the
design of follow-up studies already underway.

\section{Acknowledgements}
We thank Matthew Simmonds, Stephen Gough, Jayne Franklyn, Oliver
Brand, for sharing their AITD genetic association dataset and all AITD
patients and control subjects for participating in this study.  The
AITD UK national collection was funded by the Wellcome Trust.

Phased 1000 Genomes data was downloaded from
http://www.sph.umich.edu/csg/abecasis/MACH/download/1000G.2012-03-14.html

We would like to thank the UK Medical Research Council and Wellcome
Trust for funding the collection of DNA for the British 1958 Birth
Cohort (MRC grant G0000934, WT grant 068545/Z/02).  We acknowledge use
of data from The UK Blood Services collection of Common Controls (UKBS
collection), funded by the Wellcome Trust grant 076113/C/04/Z, by the
Wellcome Trust/JDRF grant 061858, and
by the National Institute of Health Research of England. The DNA
collection was established as part of the Wellcome Trust Case-Control
Consortium.

For additional detailed acknowledgements of the source assay design,
genotyping, data and samples, please see Cooper et al (2012).

CW is funded by the Wellcome Trust (089989).  The Diabetes and
Inflammation Laboratory is funded by the JDRF, the Wellcome Trust and
the National Institute for Health Research (NIHR) Cambridge Biomedical
Research Centre. The Cambridge Institute for Medical Research (CIMR)
is in receipt of a Wellcome Trust Strategic Award (100140).

\section{Conflict of Interest}
The authors have no conflict of interest to declare.

\renewcommand{\refname}{References}

\clearpage

\clearpage
\section{Supplementary Material}

\subsection{Simulation}

Once a ``causal variant'' SNP, $S$, was selected, control haplotypes
were sampled randomly and case haplotypes sampled conditional on the
allele carried at $S$.  For a disease model with relative risk $r$,
and given the minor (risk) allele at $S$ has frequency $\pi_0$, in
controls, the frequency in cases is
$$\pi_1 =\frac{r\pi_0}{1-\pi_0+r\pi_0}. $$ 
Therefore when sampling case
haplotypes, we over-sample haplotypes carrying the risk allele and
under sample those carrying the protective allele by using sampling
probabilities  proportional to 
\begin{equation*}
P_S =  
\begin{cases}
 \dfrac{\pi_1}{\pi_0}& \text{haplotype carries risk
  allele}\\[12pt]
\dfrac{1 - \pi_1}{1 - \pi_0} & \text{haplotype carries protective allele
  allele.} 
\end{cases}
\end{equation*}

For eQTL data, we simulated a response variable, $Y$ as a mixture of Gaussians
\[Y = \sqrt{0.7}Z + \sqrt{0.3}{X}\] where $Z$ was sampled from a
standard Gaussian and $X$ is the count of the minor allele at the
causal SNP.  Thus, $X$ would explain 30\% of the variance of $Y$, or
30\% of the simulated eQTL, independent of minor allele frequency.


\subsection{The effect size at a selected SNP}

To calculate the bias in figure \ref{fig:bias-effect}, we compared the estimated effect
size at the sampled SNP to the true effect \emph{at that SNP}, ie not
at the causal SNP.  If the causal SNP is $S$ and the selected SNP is
$T$, then the underlying relative risk  at $T$ is simplest to calculate
in a haploid system, which is equivalent to assuming Hardy Weinberg
equilibrium.  Given
$$\rho' = \rho(S,T)  \sqrt{\pi_S \pi_T (1-\pi_S) (1-\pi_T)}$$
where $\rho(S,T)$ is the correlation between $S$ and $T$, then the
expected proportion of cases in the population conditional on the
allele carried at SNP $T$ is
\begin{equation*}
  \begin{cases}
    D_1 = \dfrac{r (\pi_S\pi_T + \rho') + ((1-\pi_S)\pi_T - \rho')}{\pi_T} &
    \text{$T$ is risk allele}\\[12pt]
    D_0 = \dfrac{r (\pi_S(1-\pi_T) - \rho') + ((1-\pi_S)(1-\pi_T) +
      \rho')}{1-\pi_T} & \text{$T$ is protective allele}\\
  \end{cases}
\end{equation*}
and the relative risk is $\dfrac{D_1}{D_0}$.  For a rare disease such
as T1D, relative risks and odds ratios are approximately equal.

\subsection{Implementation of Bayesian Model Averaging}

Bayesian model averaging requires evaluating all possible multiple SNP models in each trait, and conducting colocalisation testing for each model.  We began by defining the posterior probability of model $j$ for both traits as, 
\begin{equation*}
\pi_j = \frac{\pi_j^1\pi_j^2}{\sum_k{ \pi_k^1\pi_k^2 }}
\end{equation*}
where $\pi_j^i$ is the posterior probability of model $j$ for trait
$i$ and a model $j$ indicates which SNPs are included in the model.
Even when the number of SNPs to be tested is fixed at two, the number
of possible models is $\tfrac{p!}{2!(p-2)!}$.  Whilst testing all models is feasible if computationally expensive for analysis of real data, it is impractical for simulations.  To reduce the computation burden, we first evaluated all $p$ single SNP models and indentified the set of SNPs with very low posterior probability ($\pi_j < 0.01$).  We then excluded any two SNP model containing \emph{only} SNPs from this set.  If $p_0 < p$ such SNPs were identified, this reduced the number of models to test to 
$\tfrac{p!}{2!(p-2)!} - \tfrac{p_0!}{2!(p-2)!}$.  

For the purposes of simulation, we used the profile likelihood
approach to generate a $\chi^2_1$ distributed test statistic and
averaged the resulting $p$ values, $P_j$, over the model space to calculate an
overall posterior predictive $p$ value, $\sum_jP_j\pi_j$.  For the
application to AITD, we integrated the $p$ value associated with the
$\chi^2_2$ distributed test statistic calculated assuming $\eta$ was
known over both the posterior distribution of $\eta$ given each model,
and the posterior of the model space.  The latter is formally correct,
but comptationally too expensive for simulation, and the profile
likelihood $p$ value and the posterior predictive $p$ value have been
shown to be very similar for large samples.

\clearpage
\renewcommand{\tablename}{Supplementary Table}
\renewcommand{\figurename}{Supplementary Figure}

  \begin{sidewaystable}
    \centering
    \begin{tabular}{llrrlrlrr}
\toprule
      Region & SNP & MAF & \multicolumn{2}{c}{GD} &
      \multicolumn{2}{c}{HT} & \multicolumn{2}{c}{Power, $\alpha=$} \\
\cmidrule(lr){4-5}\cmidrule(lr){6-7}\cmidrule(lr){8-9}
      &&& OR  & \multicolumn{1}{c}{p value} & OR  & \multicolumn{1}{c}{p value}  & $0.05$ & $10^{-6}$ \\
\midrule
\multicolumn{9}{l}{\emph{Associated with GD and HT in published study}}\\
1p13.2/\emph{PTPN22} & rs2476601 G$>$A & 0.096 & 1.55 & $4.03\times10^{-16}$ & 2.02 & $3.74\times 10^{-15}$ & 0.99 & 0.37\\
2q33.2/\emph{CTLA4}/\emph{ICOS} &rs11571297 G$>$A &0.493&0.72&$2.81\times10^{-23}$&0.82&$3.21\times10^{-3}$&0.99 & 0.490\\
2p25.1/\emph{TRIB2} &rs1534422 A$>$G &0.455 &1.16 &$4.69\times10^{-6}$ &1.24 & $1.64\times10^{-3}$ &0.61&0.004\\
3q27.3/3q28/\emph{LPP}&rs13093110 C$>$T &0.452 &1.18 &$8.17\times10^{-7}$ &1.20 & $7.09\times10^{-3}$ &0.70&0.008\\
6q15/\emph{BACH2}&rs72928038 G$>$A &0.177 &1.21 &$3.63\times10^{-6}$ &1.30 & $1.36\times10^{-3}$ &0.64&0.005\\
10p15.1/\emph{IL2RA} &rs706779 A$>$G &0.467&0.85&$2.27\times10^{-6}$&0.84&0.0125&0.674 & 0.007\\
11q21/-- &rs4409785 T$>$C &0.173 &1.21 & $5.37\times10^{-6}$ &1.34 &
$3.54\times10^{-4}$ &0.63&0.004\\
\\
\multicolumn{9}{l}{\emph{Associated with GD only in published study}}\\
1p36.32/\emph{TNFRSF1} &rs2843403 C$>$T &0.362 &0.84 &$7.94\times10^{-7}$ & 0.97 & 0.696 &0.69&0.007\\
1q23.1/\emph{FCRL3} &rs7522061 T$>$C &0.480&1.16&$1.08\times10^{-5}$ &1.03&0.634&0.60 & 0.004 \\
6q27/\emph{CCR6} &imm\_6\_167338101 A$>$C &0.408 &0.84 &$3.30\times10^{-7}$ &0.88 & 0.056 &0.71&0.009\\
12q12/\emph{PRICKLE1}&rs4768412 C$>$T &0.363 &1.19 &$3.30\times10^{-7}$ &1.00 & 0.949 &0.73&0.010\\
14q31.1/\emph{TSHR} &rs2300519 T$>$A &0.380&1.54&$1.34\times10^{-38}$ &0.93&0.295&1 & 0.95\\
16p11.2/\emph{ITGAM}&rs57348955 G$>$A &0.396 &0.83 &$3.76\times10^{-8}$ &0.91 & 0.188 &0.76&0.013\\
\bottomrule
    \end{tabular}
    \caption[Power to detect association with Hashimoto's Thyroiditis
    to confirmed loci for Graves' Disease]{\textbf{Power to detect
        association with Hashimoto's Thyroiditis to confirmed loci for
        Graves' Disease.} Region denotes chromosomal region and most
      likely candidate gene(s) where available [9].
      GD=Graves' Disease; HT=Hashimoto's Thyroiditis; MAF=minor allele
      frequency in controls.  }
    \label{tab:power-detect-assoc}
  \end{sidewaystable}
  
\clearpage


\begin{figure}
  \centering
  \includegraphics[width=\textwidth]{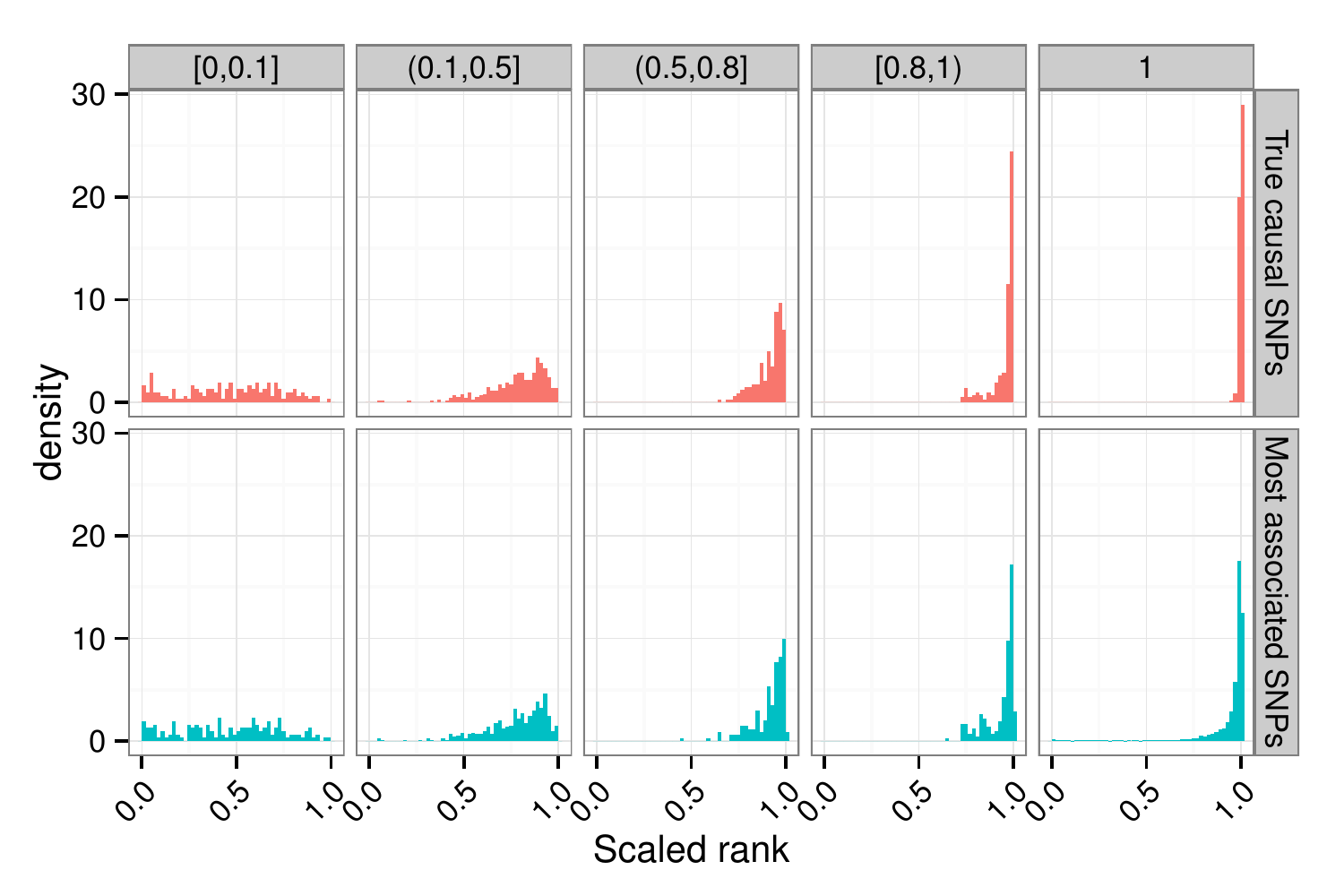}
  \caption[Distribution of Nica \emph{et.\ al}'s rank
  statistic]{\textbf{Distribution of Nica \emph{et.\ al}'s rank
      statistic.}  The statistic is evenly distributed within [0,1]
    when the LD between the causal variants is negligable, but is
    increasingly biased towards 1 as the LD increases.  Columns are
    divided by the $r^2$ between distinct causal variants, with
    $r^2=1$ indicating a shared causal variant.  The top row shows the
    optimal result that could be obtained if conditioning on the true
    causal variant were possible, the bottom row shows the effect of
    conditioning on the most associated SNP is to reduce the degree of
    skew.  Results are shown for a complete genotyping scenario, with
    a sample size of 2000 and a relative risk of 1.3.  Similar effects
    are seen under tagging or complete genotyping approaches, but the
    skew towards 1 occurs more rapidly with larger samples and effect
    sizes.}
\label{fig:nica}
\end{figure}

\end{document}